\pdfoutput=1
\RequirePackage{ifpdf}
\ifpdf 
\documentclass[pdftex]{sigma}
\else
\documentclass{sigma}
\fi

\numberwithin{equation}{section}

\newcommand{\bphi}{\boldsymbol{\phi}}
\newcommand{\bp}{\boldsymbol{\phi}}
\newcommand{\bPhi}{\boldsymbol{\Phi}}

\newcommand{\bX}{\boldsymbol{X}}

\newcommand{\bmu}{\boldsymbol{\mu}}

\begin{document}
\allowdisplaybreaks

\renewcommand{\thefootnote}{}

\newcommand{\arXivNumber}{2211.02413}

\renewcommand{\PaperNumber}{034}

\FirstPageHeading

\ShortArticleName{Stable Kink-Kink and Metastable Kink-Antikink Solutions}

\ArticleName{Stable Kink-Kink and Metastable Kink-Antikink\\ Solutions\footnote{This paper is a~contribution to the Special Issue on Topological Solitons as Particles. The~full collection is available at \href{http://www.sigma-journal.com/topological-solitons.html}{http://www.sigma-journal.com/topological-solitons.html}}}

\Author{Chris HALCROW and Egor BABAEV}
\AuthorNameForHeading{C.~Halcrow and E.~Babaev}
\Address{Department of Physics, KTH-Royal Institute of Technology, Stockholm, SE-10691 Sweden}
\Email{\href{mailto:chalcrow@kth.se}{chalcrow@kth.se}, \href{mailto:babaev@kth.se}{babaev@kth.se}}

\ArticleDates{Received February 21, 2023, in final form May 23, 2023; Published online June 01, 2023}

\Abstract{We construct and study two kink theories. One contains a static 2-kink configuration with controllable binding energy. The other contains a locally stable non-topological solution, which we call a lav\'{\i}on. The new models are 1D analogs of non-integrable systems in higher dimensions such as the Skyrme model and realistic vortex systems. To help construct the theories, we derive a simple expression for the interaction energy between two kinks.}


\Classification{35C08; 35Q51; 37K40}

\renewcommand{\thefootnote}{\arabic{footnote}}
\setcounter{footnote}{0}

\section{Introduction}

Kinks are the prototypical example of topological solitons in field theories, and model domain walls in numerous important physical systems. Many kink properties can be calculated: their masses, interactions \cite{manton1979}, dynamics \cite{Sugiyama1979} and quantum loop corrections \cite{Cahill1976, Evslin2019}. Remarkably, many of these properties can be described analytically, owing to the simplicity of the one-dimensional theories. Kink theories are then an excellent place to experiment with difficult concepts which are important in higher-dimensional soliton theories.

However, there are several higher-dimensional theories whose kink analogs are unknown, or have been minimally studied. We're ultimately interested in non-integrable theories in higher dimensions, such as realistic vortex models and nuclear skyrmions. These share two features that we will try to replicate in simple kink-theories: that an $n$-soliton can be split into $n$ 1-solitons and that the solutions have no additional `surprising' symmetries. More technically, the kink moduli space should be equal to the symmetry group of the model and the field theory must contain a saddle point equal to $n$ infinitely separated 1-solitons. Models with these features are analogs of the non-integrable higher-dimensional models. A one-dimensional system with these properties was discovered in the supersymmetric Wess--Zumino model \cite{portugues2002intersoliton}, whose kink dynamics was studied in \cite{alonso2021kink}. Kink models with extra symmetries (such as the MSTB \cite{montonen1976solitons, sarker1976solitary} or BNRT \cite{bazeia1997soliton, shifman1998degenerate}) are analogs of integrable soliton theories such as instantons, lumps and critically coupled vortices and monopoles.

Why is it worth examining kink theories with new features? Consider the set of new theories proposed in \cite{adam2019}, which couple a kink to a carefully chosen defect. Here, the authors constructed a kink model whose spectral structure depends on the position of the kink, mimicking a generic feature of soliton systems in higher dimensions for the first time. Since the kink is one-dimensional, the authors were able to study the problem in great detail. The analysis revealed a new theoretical feature, a spectral wall, which has now been shown to exist in a broad range of one-dimensional theories \cite{adam2019thick, adam2021}, affects quantum interactions~\cite{evslin2022} and should exist in higher dimensions too. Further, the authors used insights from the special theories to construct a successful collective coordinate approximation to $\Phi^4$ kink-antikink dynamics for the first time~\cite{manton2021}. So the study of a kink theory with a new property led to the discovery of a new theoretical concept and a new approximation for an old model.

The main feature we study are stable, isolated multi-kink configurations. These include a~stable kink-kink pair and a metastable kink-antikink pair. In generic scalar theories, kink-kink configurations repel while kink-antikink configurations attract and annihilate. But in higher-dimensional models, bound multisoliton states are often the most interesting part of the theory. Multiskyrmions in 3D model finite nuclei, but there is much debate about their quantisation and binding energies. Hence constructing an appropriate toy model where calculations can be done in detail may provide a road-map for more difficult problems. Recently multiskyrmion solutions attracted interest in condensed matter, such as non-monotonically interacting vortex clusters in type-1.5 superconductors \cite{babaev2005,Carlstrom2011}. In that context it is interesting to have simpler, more analytically tractable one-dimensional solutions.

One of the theories supports a locally stable kink-antikink configuration. These solutions model domain wall pairs. Pairs that can form robust clusters or stable domain-antidomain wall solutions should have potential advantages in domain wall racetrack memory \cite{Parkin2008}, overcoming the usual disadvantage that pairs of walls tend to annihilate. Our new solution is locally stable, meaning that materials supporting these configurations will be good candidates for racetrack memory devices if realized in realistic physical systems. In a forthcoming paper, we report the existence of these stable wall-wall pairs in ferroelectrics, providing further theoretical imperative for the development of ferroelectric domain wall racetrack memory \cite{Catalan2012, Sharma2017}.

Many soliton theories (including kink theories \cite{rajaraman1979}) support sphalerons \cite{Manton1983, Manton2019}, static saddle points which are sometimes constructed from a soliton-antisoliton pair. Our configuration is different, a true minimum of the theory. Its discovery begs the question: do similar solutions exist in other soliton theories? Numerical evidence of a stable vortex-antivortex excitation in a spin imbalanced superfluid was presented in \cite{Barkman2020}, though it required the nontrivial background of so-called Fulde--Ferrell state and higher-order gradient terms in the theory. Hence it is quite a different situation from the one considered here. There is also a history of interesting kink-antikink configurations. Rajaramen created a kink-antikink solution using an ingenious ansatz~\cite{rajaraman1979}, but this is not isolated since there is a one-dimensional family of solutions (controlled by their parameter $f$). This might sound like a small difference, but it means that Rajaraman's solutions have a zero-energy mode which can be used to flow the field to the vacuum using no energy. Hence the physics is very different than what we present: where the kink-antikink solution has roughly twice the energy of its constituent parts. Dynamically stable soliton-antisoliton configurations have been found, such as internally rotating domain walls \cite{eto2008domain}. The model presented in \cite{halavanau2012} may also support stable kink-antikink solutions. Some models have interesting static ``non-topological'' excitations whose fields look like ours, but is unstable \cite{alonso2009}. The most similar configuration is described in \cite{alonso2020non}. Their solution cannot be split into an infinite separated kink and antikink, since the theory has only one vacuum. Hence their configuration is a stable analog of the electroweak sphaleron~\cite{Manton1983} where ours is a stable analog of the monopole-antimonopole sphaleron~\cite{taubes1982existence}.\looseness=-1

This paper is organised as follows. In the next section we will derive a simple formula for the interaction of two well separated kinks. We will then use this formula to help construct a theory with a static, stable kink-kink solution in Section~\ref{section3} and a theory with a stable kink-antikink solution in Section~\ref{section4}.

\section{Interactions}\label{section2}

In this section we will call any solution linking two vacua a kink; choosing not to distinguish between kinks and antikinks. The interaction between kinks is usually calculated using an argument concerning the force felt by one kink from another, due to Manton \cite{manton1979}. Here, we present an alternative method by directly evaluating the energy of two well separated configurations. Such a method has been developed for skyrmions \cite{barton2022stability, feist2012, Schroers1993}. An advantage of this method is that it does not rely on the concepts of force and momentum. Hence the results apply to theories with no second order time evolution, such as domain walls in condensed matter systems.

Consider a multicomponent kink theory with energy density
\[
\mathcal{E} = \tfrac{1}{2}\partial_x\Phi_a G_{ab} \partial_x\Phi_b + V(\Phi) ,
\]
where we take the components of $G$ to be constant. The theory has Euler--Lagrange equations
\begin{equation} \label{eq:ELkink}
	G_{ab}\partial_x^2 \Phi_b - \partial_a V(\Phi) = 0 .
\end{equation}
The vacua of the theory, which we'll denote $\Phi^v$, satisfy $\partial_a V(\Phi^v) = 0$. Consider small fluctuations around the vacua
\[
\Phi_a(x) = \Phi^v_a + \phi_a(x) .
\]
The Euler--Lagrange equations for the fluctuations are
\begin{equation*} 
	\partial_x^2 \phi_a - G^{-1}_{ab} \partial_b\partial_c V\left( \Phi^v \right)\phi_c = 0 ,
\end{equation*}
which have solution
\[
\bp = \sum_n \bmu_n {\rm e}^{-\sqrt{\lambda_n} x} ,
\]
where $\lambda_n$ and $\bmu_n$ are the eigenvalues and eigenvectors of $G^{-1}_{ab}\partial_b\partial_c V(\Phi^v)$. For simplicity, we'll assume that there is no zero-eigenvector. The leading behaviour is described by the eigenvector with the smallest eigenvalue.

Now consider two kinks at positions $\pm X$ with $|X| \gg 1$. These are static solutions to the equations of motion \eqref{eq:ELkink}. Denote the kinks as $\Phi^{\pm X}$ with tails $\bphi^{\pm X}$. The two kinks link three vacua: $\Phi^{v_{-\infty}}$, $\Phi^{v_0}$ and $\Phi^{v_{\infty}}$. For continuity, the walls must share the same central vacuum $\Phi^{v_0}$. We can then combine the walls into one field,
\[
\Phi(x) = \Phi^{-X}(x) + \Phi^X(x) - \Phi^{v_0} .
\]
This is a solution of the Euler--Lagrange equations in the limit $|\bX| \to \infty$. Far from the wall centers, the above asymptotic analysis applies so we can write
\begin{gather*}
	\Phi(x)  \approx \begin{cases} \Phi^{-X}(x) + \bp^X(x) - \Phi^{v_0}& \text{for $x<0$}, \\
		\bp^{-X}(x) + \Phi^X(x) - \Phi^{v_0}& \text{for $x>0$}. \end{cases}
\end{gather*}
Using this approximation, we can evaluate the energy of the configuration as a Taylor series in~$\phi$. To do so, we split the real line in half so that
\begin{gather*}
	E(\Phi) \approx \int_{0}^\infty \Big(\tfrac{1}{2}G_{ab} \partial_x \Phi^{X}_a \partial_x \Phi^X_b + V\big(\Phi^X\big)
	+ G_{ab}\partial_x \Phi^X_a \partial_x \phi^{-X}_b + \phi^{-X}_a \partial_a V\big( \Phi^X \big) \Big)\,{\rm d}x   \\
\hphantom{E(\Phi) \approx}{} +\int_{-\infty}^0 \Big(\tfrac{1}{2}G_{ab} \partial_x \Phi^{-X}_a \partial_x \Phi^{-X}_b + V\big(\Phi^{-X}\big)
	+ G_{ab}\partial_x \Phi^{-X}_a \partial_x \phi^{X}_b + \phi^X_a \partial_a V\big( \Phi^{-X} \big)\Big)  \,{\rm d}x .
\end{gather*}
We can simplify the order $O\big(\phi^0\big)$ terms since these are the energies of the $\Phi^X$ and $\Phi^{-X}$ kinks (up to exponentially small corrections). We then simplify the $O(\phi)$ term by integrating by parts. The total energy is then
\begin{gather*}
E(\Phi) = E\big(\Phi^X\big) + E\big(\Phi^{-X}\big) + \big[G_{ab} \partial_x \Phi^X_b \phi^{-X}\big]_0^\infty + \big[G_{ab} \partial_x \Phi^{-X} \phi^{X}\big]_{-\infty}^0 \\
\hphantom{E(\Phi) =}{}
+\int_{0}^\infty \phi^{-X}_a \left( -G_{ab} \partial^2_x \Phi^{X}_b + \partial_a V\big( \Phi^X \right) \big) \,{\rm d}x\\
\hphantom{E(\Phi) =}{}
 + \int_{-\infty}^0
		\phi^{X} \bigl( -G_{ab}\partial^2_x \Phi^{-X}_b +\partial_a V\big( \Phi^{-X} \big) \bigr) \,{\rm d}x .
\end{gather*}
The terms in the integrals are equal to zero due to the equations of motion~\eqref{eq:ELkink}. Hence, only the boundary term survives. Then, since $X\gg 0$, we can use $\Phi^{\pm X} \approx \phi^{ \pm X}$ at $x=0$. Finally, we get a simple expression for the interaction energy
\begin{equation} \label{eq:Eint}
	E^\text{int}(\Phi) = G_{ab}\big( \partial_x \phi_a^{-X} \phi_b^X - \partial_x \phi_a^X \phi_b^{-X} \big) \big\rvert_{x=0} .
\end{equation}
The interaction energy only depends on the field tails at the point between the two kinks. Note that this result is true for constant~$G$. A similar, modified result will hold if $G$ is a function of space or the fields themselves.

\subsection[An example: Phi\^6 theory]{An example: $\boldsymbol{\Phi^6}$ theory}\label{section2.1}

As a test of the simple expression \eqref{eq:Eint}, consider $\Phi^6$ theory. Here,
\[
	G = 1 , \qquad V_6(\Phi) = \tfrac{1}{2}\Phi^2\big(1-\Phi^2\big)^2 .
\]
There are three vacua: $\Phi = -1, 0$ and $1$. A kink connects vacua from left to right in an ascending order while an antikink does so in descending order. We can label a kink or antikink by the two vacua they connect as $\Phi_{(v_1,v_2)}$. Explicit formula for the solutions are known and given by
\begin{align} \label{eq:kink}
\Phi^K_{(0,1)}(x) = -\Phi^K_{(-1,0)}(-x) = \Phi^{\bar{K}}_{(1,0)}(-x) = -\Phi^{\bar{K}}_{(0,-1)}(x) = \frac{1}{\sqrt{1+3{\rm e}^{-2x}}} .
\end{align}
The kinks have a zero mode arising from translational symmetry, meaning that any shifted configuration (such as $\Phi^K_{(0,1)}(x-X)$) is also a static solution. The moduli $X$ is interpreted as the kink position. Another definition of position is where the field takes the value between the vacua it connects. In this case when $|\Phi| = 1/2$, which also occurs at $x=X$.

Now consider two scenarios: the interaction of two kinks, and the interaction of a kink and an antikink. A kink-kink pair with positions $-X$ and $X$ are approximated by
\[
	\Phi_{(-1,0)}(x+X) + \Phi_{(0,1)}(x-X) .
\]
To evaluate the interaction energy we calculate the tails at $x=0$. We can write down the kink tails directly from \eqref{eq:kink} or by evaluating $V_6''(0) = 4$. Either way, the tails are given by
\begin{gather*}
	\phi^{-X}(x) =\phi_{(-1,0)}(x+X)  = \tfrac{3}{2}{\rm e}^{ -2(x+X) }, \qquad -X \ll x, \\
	\phi^X=(x) = \phi_{(0,1)}(x-X)  = -\tfrac{3}{2}{\rm e}^{ 2(x-X) }, \qquad x \ll X .
\end{gather*}
The interaction energy is then
\begin{gather*}
	E^\text{int}= \frac{9}{4}\partial_x \big({\rm e}^{ -2(x+X) }\big)\bigl(-{\rm e}^{ 2(x-X) }\bigr) - \frac{9}{4}\partial_x \bigl(-{\rm e}^{ 2(x-X)}\bigr)\big({\rm e}^{ -2(x+X) }\big) \big\rvert_{x=0} = 9 {\rm e}^{-4X} .
\end{gather*}
The interaction is positive and the kinks can lower their energy by increasing $X$. Hence the two objects repel and there is no static kink-kink configuration.

The calculation for a kink-antikink pair is identical, with the replacement $\Phi_{(0,1)} \to \Phi_{(0,-1)}$. This changes the sign of the tail $\phi^{X}$. With this update, the interaction energy is
\begin{gather*}
	E^\text{int} = \frac{9}{4}\partial_x \big({\rm e}^{ -2(x+X) }\big)\big({\rm e}^{ 2(x-X) }\big) - \frac{9}{4}\partial_x \big({\rm e}^{ 2(x-X)} \big)\big({\rm e}^{ -2(x+X) }\big) \big\rvert_{x=0} = -9 {\rm e}^{-4X}.
\end{gather*}
The interaction energy is negative and can be lowered decreasing $X$. Hence the kink and antikink attract and will, eventually, annihilate into the vacuum.

The result here is generic in one-component theories: a kink repels another kink but attracts an antikink. The attraction or repulsion hinges on the signs and derivatives of the tails, as seen above. Hence we can intuit whether solitons attract or repel by studying the graph of the fields. If the fields ``look like" the tail of two kinks, they will repel while if they ``look like'' the tail of a~kink-antikink, they will attract. This simple intuition may break in more complicated systems, but will prove useful for the rest of this paper.

\section{A theory with a stable 2-kink configuration}\label{section3}

We begin by defining the naming convention for our kinks. There is no consistent naming terminology for multikinks, and any that is chosen is based somewhat on interpretation. In this section, we will study a theory whose vacua have a well defined order (in our case, their $\Phi_1$ value). In this case, we call any configuration joining the $i^\text{th}$ and $(i+n)^\text{th}$ vacua an $n$-kink. Since $n$ only depend on the boundary data, it is an implicit property of the kink (although depends crucially on the orderability of the vacua). This terminology is consistent with what is used for sine-Gordon theory. Assuming monoticity (which only holds in simple models) we may define the position(s) of a kink as the preimage of the points halfway between two vacua. In this way, an $n$-kink has $n$ positions, which may then be interpreted as the positions of its constituent $n$ 1-kinks. This description is reminiscent of using zeros of the Higgs field as vortex positions and preimages of the antivacuum as skyrmions positions.

\begin{figure}[t]\centering
\includegraphics[width=1.0\columnwidth]{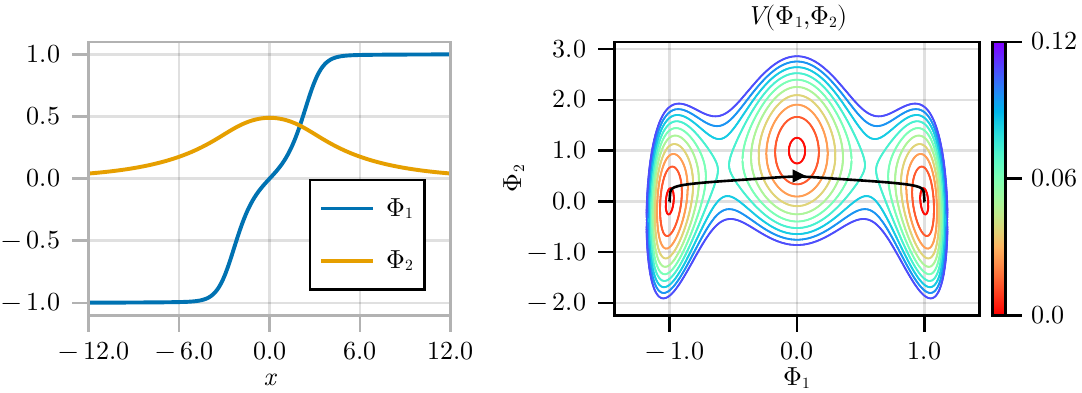}
\caption{The energy minimising 2-kink of \eqref{eq:potkk} with $\mu=1/4$, $m=1$. We plot the fields as functions of $x$ (left) and the solution in potential space (right).}		\label{fig:soln}
\end{figure}

Consider a two-component theory with a $\Phi^6$ theory embedded in the first component. We will construct a 2-kink solution joining its three vacua: $-1$, $0$ and $1$. As we saw previously, the interaction energy is determined by the field behaviour in the central vacuum, $\phi = 0$. The kinks in the first component will repel, so we would like the fields in the second component to attract. Hence their tails should mimic the tails of a kink-antikink configuration. One way of achieving this is to construct a potential whose central vacuum has a non-zero second component. Then the initial 2-kink configuration will join $(1,0)$, $(0,m)$ and $(-1,0)$. Assuming monoticity between the vacua, the second field component will act as desired. A configuration which matches our requirements is seen later in the text, in Figure~\ref{fig:soln} (left).

So, we seek a potential $V(\Phi_1,\Phi_2)$ with three vacua $(1,0)$, $(0, m)$ and $(-1,0)$. For single kinks to exist, the three vacua must have the same energy. A simple example of a potential with these features is
\begin{equation} \label{eq:potkk}
	V(\Phi_1,\Phi_2) = \tfrac{1}{2}\Phi_1^2\big(1-\Phi_1^2\big)^2 + \frac{\mu^2}{2}\left(1-\Phi_1^2 - \frac{\Phi_2}{m}\right)^2 .
\end{equation}
We'll consider the theory consisting of this potential and the simplest metric $G_{ab} = \delta_{ab}$. A similar theory was studied in~\cite{alonso2021kink}, but our interpretation of the configurations is novel. Like the usual $\Phi^6$ theory, there are two kinks: one connecting $(-1,0)$ to $(0,m)$ and the other connecting $(0,m)$ to $(1,0)$. We label the kink joining two vacua as $\Phi^{\boldsymbol{v}_1 \to \boldsymbol{v}_2}$. Again, we define the position to be the point where the field's first component is equal to $1/2$. The kinks at position $0$ are related by a simple transformation
\[
\big(\Phi_1^{(-1,0)\to (0,m)}(x), \Phi_2^{(-1,0)\to (0,m)}(x) \big) = \bigl(-\Phi_1^{(0,m)\to (1,0)}(-x), \Phi_2^{(0,m)\to (1,0)}(-x) \bigr) .
\]

There is a configuration joining the vacua $(-1,0)$ to $(1, 0)$, which we call a 2-kink. Our interpretation is based on the fact that one can smoothly deform the 2-kink into two infinitely separated 1-kinks, which are the static solutions discussed above. Now consider the interactions of these widely separated 1-kinks. We position the first kink at $-X$ and the second at $X$. The asymptotic interaction is controlled by the Hessian at $(0,m)$:
\[
	\frac{\partial^2 V}{\partial \phi_a \partial \phi_b} \bigg\rvert_{\Phi = (0, m)} = \begin{pmatrix} 1 & 0 \\ 0 & \mu^2/m^2 \end{pmatrix} .
\]
The kink tails are then given by
\begin{gather}
\big(\phi_1^{-X}, \phi_2^{-X} \big)= \bigl(-a {\rm e}^{-(x+X)}, -b {\rm e}^{-\mu/m(x+X)} \bigr), \qquad x \gg -X,\nonumber \\
\big(\phi_1^{X}, \phi_2^{X} \big) = \big(a {\rm e}^{(x-X)}, -b {\rm e}^{\mu/m(x-X)} \big), \qquad x \ll X ,\label{eq:phi6int}
\end{gather}
where the signs are chosen so that $a,b>0$. Using \eqref{eq:Eint}, the interaction energy is
\[
	E^{\text{int}}(X) = a^2 {\rm e}^{-2X} - b^2 \mu/m {\rm e}^{-2\mu X/m} .
\]
The leading behavior depends on the size of $\mu/m$. For $\mu/m<1$, the second component will dominate, providing long range attraction. At shorter range the first component also becomes important and will provide a repulsive interaction. The kinks cannot continue to attract forever and so at some point the repulsion balances the attraction and there is a minimum in the interaction energy.

Now let us consider an explicit example, with $\mu=1/4$ and $m=1$. The 1-kink has energy $E_1 = 0.299$, and the fields fall away asymptotically with $a=1$ and $b=1/2$. We can construct the stable 2-kink solution by forming initial data which respects its topology, such as
\[
(\phi_1, \phi_2) = \big( (\tanh(x-X) + \tanh(x+X))/2 , m\exp\bigl(-x^2\bigr) \big) ,
\]
and applying a gradient flow numerically. We use the Julia package DifferentialEquations.jl to apply the gradient flow on a grid of 3600 points with lattice spacing~0.01 and fourth-order accuracy derivatives~\cite{rackauckas2017differentialequations} We find the energy-minimising solution and it is plotted in Figure~\ref{fig:soln}, alongside a contour plot of the potential energy~\eqref{eq:potkk}. All plots were made using the Julia package Makie.jl~\cite{DanischKrumbiegel2021}. The 2-kink has energy $E_2 = 0.579$. The percentage binding energy per soliton is
\begin{equation} \label{eq:bind}
	E_\text{bind} = \frac{(2 E_1 - E_2)}{2E_1} .
\end{equation}
The binding energy \% per soliton in this example is then $E_\text{bind} = 3.98\%$.

\begin{figure}[t]\centering
	\includegraphics[width=\columnwidth]{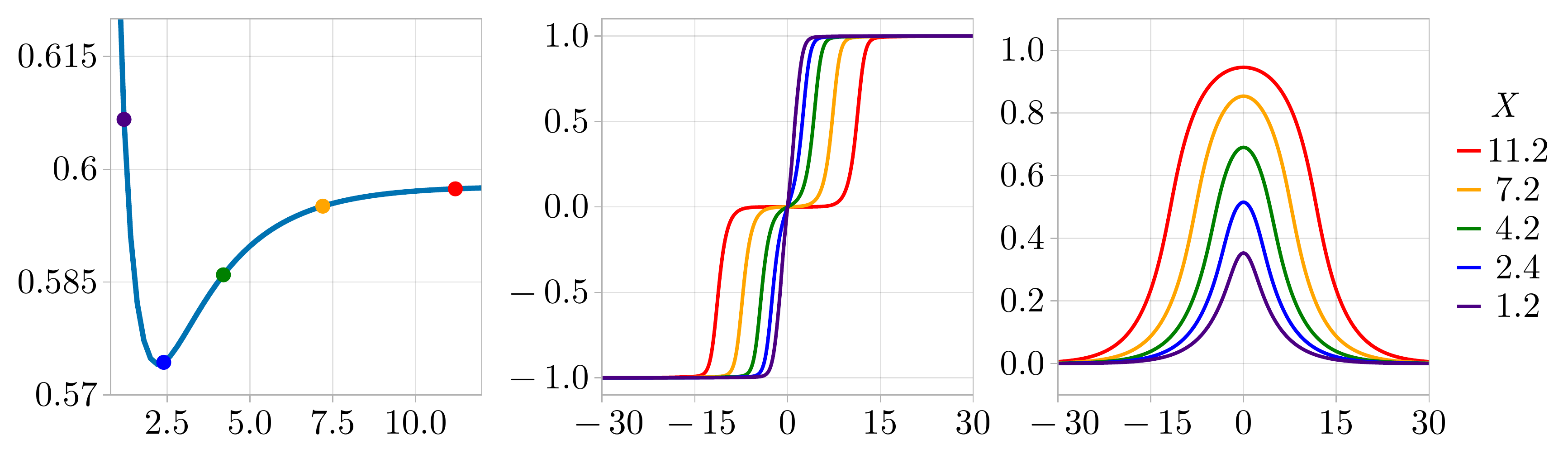}
		\caption{The energy of two kinks as a function of separation $X$ (left). The first and second field components, $\Phi_1$ and $\Phi_2$ for a selection of configurations (middle, right).}\label{fig:Ekk}
\end{figure}

The interaction potential of the two kinks can be investigated by constructing configurations with arbitrary separation. We do this by pinning: adding a constraint to fix the position of the kinks so that $\Phi_1(X) = \Phi_1(-X) = 1/2$. We then apply gradient flow again, with the constraint in place, to find the minimum energy configuration with separation $2X$. We do this and plot the energy of the configuration as a function of $X$, and some representative configurations in Figure \ref{fig:Ekk}. We know the form of the asymptotic energy from~\eqref{eq:phi6int}, which is also plotted. The potential has a short repulsive core and a long attractive tail, reminiscent of the central nucleon-nucleon potential. Another set of models with attractive long-range forces balanced by short-range repulsion are multicomponent Ginzburg--Landau theories, which describe type-1.5 superconductors~\cite{babaev2005}.

\begin{figure}[t]\centering
\includegraphics[width=\columnwidth]{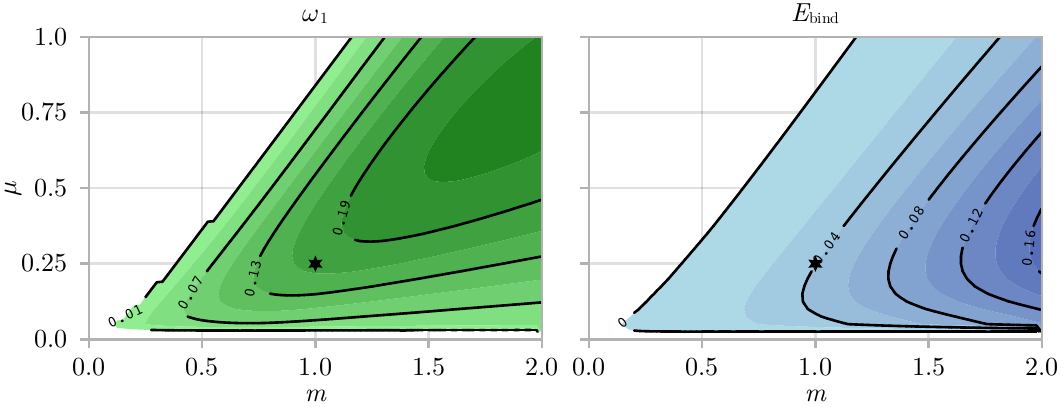}
\caption{The binding energy per soliton, as a percentage of the 1-soliton mass. The theory studied in the paper, with $\mu=1/4$, $m=1$ is highlighted by a star.}		\label{fig:bind}
\end{figure}

We can also investigate the stability of the 2-kink by calculating the linearised fluctuations around the solution $\bPhi^0$. A normal mode $\epsilon_a(x)$ with frequency $\omega$ satisfies
\begin{equation} \label{eq:normal}
 -\partial_x^2 \epsilon_a + G_{ab}^{-1}\frac{ \partial^2 V}{\partial \Phi_b \partial \Phi_c }\Big\rvert_{\bPhi^0} \epsilon_c = \omega^2 \epsilon_c .
\end{equation}
The modes and their frequencies are found by calculating the eigenvectors and eigenvalues of the discritized matrix corresponding to the operator \eqref{eq:normal}. The matrix is symmetric and sparse, meaning that even large systems can easily be evaluated. The solution in Figure \ref{fig:soln} has two normal modes. The first has frequency $\omega_0=0$ and corresponds to translations of the solution. The second has frequency $\omega_1 = 0.18$ and corresponds to the two kinks moving towards and away from each other. This is the linearised version of the mode seen in Figure \ref{fig:Ekk}. The existence of this mode, with positive frequency, guarantees that the 2-kink is stable and isolated from other non-trivial solutions. So, it is interesting to study this mode for a variety of parameters. We do so, and plot the frequency $\omega_1$ as a function of $\mu$ and $m$ in Figure~\ref{fig:bind} (left). We see that $\omega_1$ is maximal along an approximate line $m \approx 3\mu/8$. The numerical problem becomes difficult at small $\mu$ as the second component soliton has a very long tail, and on the line $\omega_1=0$ where the 2-kink stops being a minimum.

The binding energy of the 2-kink can be adjusted using $\mu$ and $m$. We calculate the binding energy \% per soliton \eqref{eq:bind} for a variety of parameters and plot the results in Figure~\ref{fig:bind}. Like the frequency, the binding energy is maximal approximately along a line, but now with a much shallower gradient, $m \approx 4\mu$. Using this new model, one could study questions about classical and quantum binding energies. The binding energy vanishes at the same point as $\omega_1$ does. This makes sense as when $\omega_1=0$ it costs no energy to pull the two kinks apart. This is the line of marginal stability, discussed in \cite{alonso2021kink, portugues2002intersoliton}. Otherwise, the normal mode frequency and binding energy functions are not tightly correlated.

\section{A theory with a metastable kink-antikink configuration}\label{section4}

In the 2-kink case, we built a theory with a long range attractive interaction and relied on the topology to ensure there was short range repulsion. In the kink-antikink case we cannot do this: the kink-antikink is in the topologically trivial sector and can collapse into the vacuum. As such, we need to build a slightly more complicated theory to support a stable kink-antikink configuration.

\begin{figure}[t]\centering
\includegraphics[width=1.0\columnwidth]{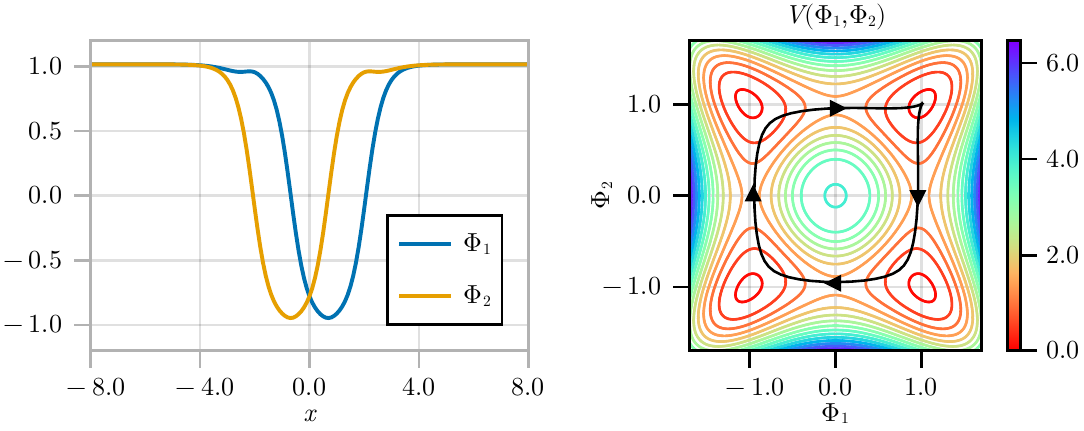}
\caption{The metastable kink-antikink solution, a lav\'{\i}on, of the theory~\eqref{eq:potkak} with $\mu_1 = 1/2$, $\mu_2 = 3$.}\label{fig:kakmin}
\end{figure}

Our idea arises by thinking about how kinks and antikinks annihilate in a multicomponent theory. Consider two weakly coupled $\Phi^4$ theories, each with vacua $\pm 1$. This gives a two-component theory with four approximate vacua: $(1,1)$, $(1,-1)$, $(-1,-1)$ and $(-1,1)$. There is a configuration which joins these four vacua in that order, then returns to~$(1,1)$. An example can be seen later in the text, in Figure~\ref{fig:kakmin}. In $(\Phi_1,\Phi_2)$ space, the field traces a square-like path, encircling the origin. For the fields to annihilate to the vacuum, the path must shrink to a~point, passing through the origin. Hence, if there is an energy cost to pass through $(\Phi_1, \Phi_2) = (0,0)$, this configuration could be stable.

In this example, the vacua are not ordered and so we cannot consistently define an $n$-kink. But for the remainder of this section, we focus on the configuration described above so we can use a naming convention which suits this study. We will call a~configuration joining~$(1,1)$ to~$(-1,-1)$ a kink and a configuration joining~$(-1,-1)$ back to~$(1,1)$ an antikink. In the theory we develop, the kink and antikink can be widely separated and hence we can apply our asymptotic analysis from Section~\ref{section2} to understand the forces between them.

Our aim is to build a two-component $\Phi^4$ theory with a long range attraction and a large energy cost to pass through~$(0,0)$. One possible example is given by $G_{ab} = \delta_{ab}$ and
\begin{gather} \label{eq:potkak}
  V(\Phi_1,\Phi_2) = \tfrac{1}{2}\big(1-\Phi_1^2\big)^2 + \tfrac{1}{2}\big(1-\Phi_2^2\big)^2 + \mu_1 \big(\Phi_1^2 - \Phi_2^2\big)^2 + \mu_2 \operatorname{sech}\big( \mu_2 \big( \Phi_1^2 + \Phi_2^2 \big) \big) .
\end{gather}
The minima satisfy $|\Phi_1| = |\Phi_2| \approx 1$. The long range interaction of a kink and antikink is determined by the Hessian at $(\Phi_1,\Phi_2) \approx (-1,-1)$, which has eigenvectors $(1,1)$ and $(-1,-1)$. The first eigenvalue is independent of $\mu_1$ while the second increases with $\mu_1$. So if we take a~large enough~$\mu_1$, the eigenvector $(1,1)$ dominates. By sketching the field $\Phi_1 + \Phi_2$ we see that this looks like a single kink-antikink. Our intuition tells us that these attract, and the interaction energy~\eqref{eq:Eint} confirms this. So, there is long range attraction provided $\mu_1>\mu_\text{critc}$. This analysis can be tightened by considering an improved vacua $|\Phi_1| = |\Phi_2| = 1 + \epsilon$ with $\epsilon$ small. There is then an attractive long range force provided $\mu_1$ is above some small critical threshold which depends on~$\mu_2$. Overall, there is a long-range attraction by a short-range repulsion due to the potential barrier at $\boldsymbol{\Phi} = \boldsymbol{0}$.

The asymptotic analysis can be tested numerically by trying to find a stable kink-antikink. We generate a kink and an antikink, place them near one another and allow the configuration to relax using gradient flow. The final solution, with $\mu_1 = 1/2$, $\mu_2 = 3$, is shown in Figure~\ref{fig:kakmin}. Unlike a sphaleron, which is a saddle point configuration, ours is a true local minimum of the theory. The sphaleron is named after its fallible character (``$\sigma \phi \alpha \lambda \epsilon \rho \acute{o} \sigma $'' means fallible, or `ready to fall' in Greek). In contrast, our configuration is stable. As such, we name the new solution a lav\'{\i}on (``$\lambda \alpha \beta \acute{\eta}$'' means grip in Greek).

Note that the two fields in Figure~\ref{fig:kakmin} have a knotted structure. The field in each component attract but to reach one another, they must cross the kink or antikink in the other component. To do so would involve passing the point $(0,0)$ which is energetically costly. So there is a stable local minimum. The knotting argument hints at a topological interpretation and there is one as follows: since $(\Phi_1,\Phi_2)= (0,0)$ is disfavoured, we can remove it from target space. The plane with a point removed has fundamental group~$\mathbb{Z}$. Hence configurations can be labelled by an integer. This topological degree counts the number of times the field winds around the origin. In reality, there is only an energy cost to reaching this point, so the topological charge is protected by an energy barrier rather than the fundamental topology of the system. We may say that the lav\'{\i}on has an additional effective topological degree.

We can check the linear stability of the lav\'{\i}on by calculating the normal modes defined in~\eqref{eq:normal}. The lav\'{\i}on has four normal modes with freqencies $\omega = 0, 0.27$, $0.52$ and $0.72$. They correspond to: translation; the kink and antikink moving towards and away from each other; one component contracting while the other expands; one component translates to the left while the other translates to the right. It is interesting that such a simple system has such a complicated mode structure.
We study the nonlinear stability of the lav\'{\i}on using a string method. Here, we build a string of configurations interpolated from the lav\'{\i}on $\phi^{l}(x)$ to the vacuum $\phi^v$:
\[
	\phi(s,x) = s \phi^v + (1-s)\phi^{l}(x), \qquad s\in [0,1] .
\]
A gradient flow is applied to the entire chain of configurations, perpendicular to the direction of the string. That is, we solve
\[
	\partial_\tau \Phi(s,x) = - \left( \frac{\delta V}{\delta \Phi} - k^{-1} \bigg \langle \frac{\delta V}{\delta \Phi} , \partial_s \Phi(s,x) \bigg\rangle \partial_s \Phi(s,x) \right) ,
\]
with $k = \langle \partial_s \Phi(s,x) , \partial_s \Phi(s,x) \rangle$ and the usual $L^2$ inner product
\[
	\langle \Phi^{(1)}, \Phi^{(2)} \rangle = \int_{-\infty}^\infty \Phi^{(1)}_a(x) \Phi^{(2)}_a(x)\,{\rm d}x .
\]
 The flow reduces the energy of the string, but does not allow points to move towards one another in field space. The final result is a low energy path in configuration space which joins the vacuum and lav\'{\i}on. Since both of these are minima the path must pass through a saddle point. The unstable mode of the saddle is in the direction of the string.

\begin{figure}[t]\centering
\includegraphics[width=1.0\columnwidth]{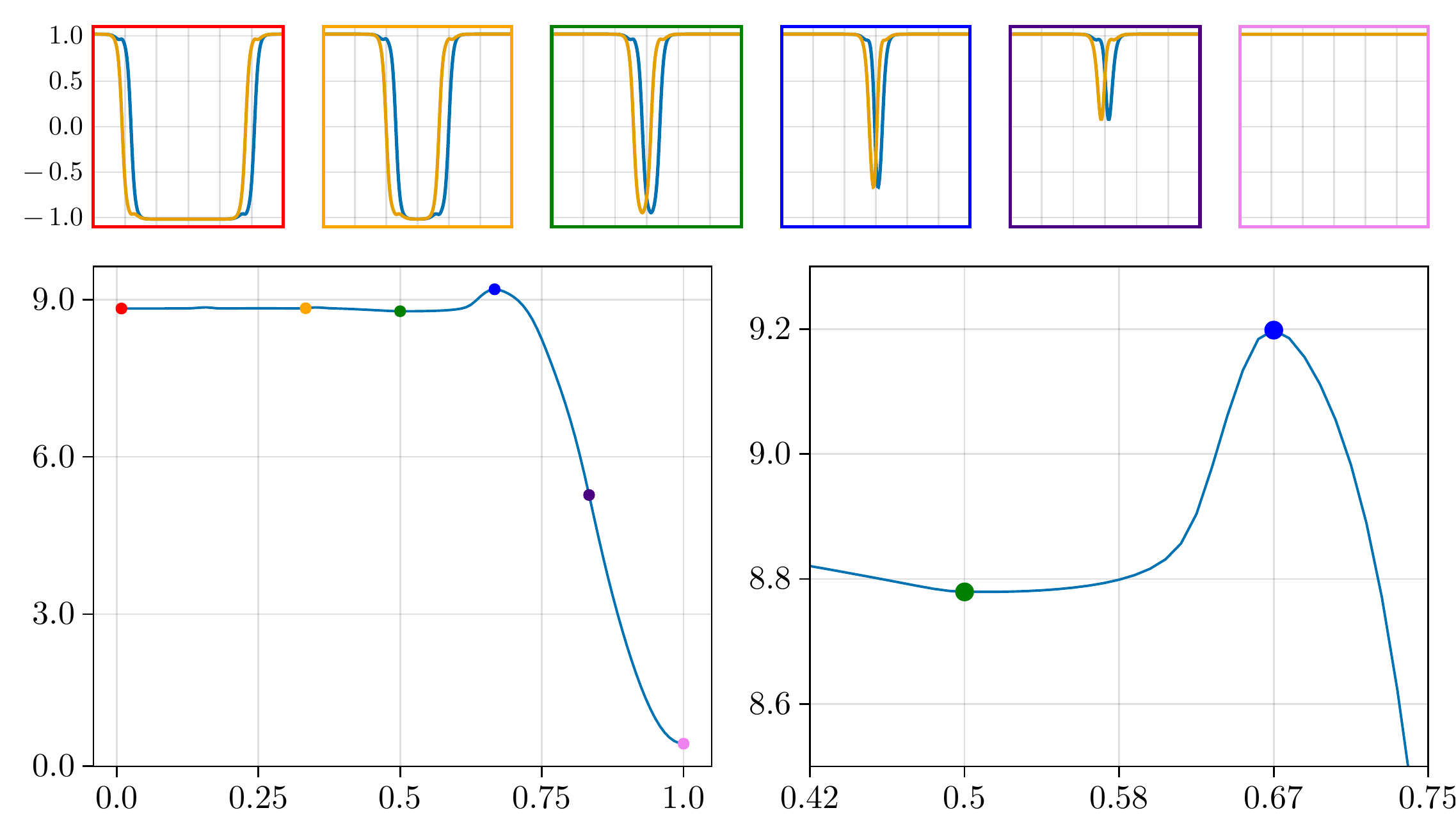}
\caption{A string of kink-antikink configurations for the model \eqref{eq:potkak} with $\mu_1=1/2$, $\mu_2=3$. The energy of many configurations are plotted as a function of distance along the string. We plot the energy function near the minima and saddle point on the right. Seven individual configurations are plotted, including the energy minimiser (yellow) and the saddle point (green). The axes for the field plots are the same as those shown in Figure~\ref{fig:kakmin}.} \label{fig:kak}
\end{figure}

We make two strings of configurations: one which joins the lav\'{\i}on to the vacuum and one which joins it to a widely separated kink-antikink. The energies of the configurations in the string, and some configurations are plotted in Figure~\ref{fig:kak}. We see two features explicitly: that the lav\'{\i}on is a local minimum and that the saddle point joining it to the vacuum is near the point of topological collapse.

There is a similar configuration studied in \cite{alonso2020non}, which we would also call a lav\'{\i}on. In that model there is only one vacuum and so the configuration cannot be pulled apart into an infinitely separated kink and antikink, and there is no low energy mode like the one studied in Figure~\ref{fig:kak}. Theirs is a~stable analog of the electroweak sphaleron \cite{manton1979}; ours is a~stable analog of the monopole-antimonopole sphalerons~\cite{taubes1982existence}.

\section{Conclusion and further work}\label{section5}

In this paper we have studied two kink theories, each with a novel feature. The first has a~stable isolated 2-kink solution, with an adjustable binding energy. This serves as a toy model for many higher-dimensional non-integrable solitons, including skyrmions. The theory is a good place to probe the connections between classical and quantum binding energies. Quantising the parameter describing the kink separation would be the 1D equivalent of studying vibrational quantisation for skyrmions~\cite{Halcrow2015}. We could also study one- (or two-~\cite{Evslin:2021loops})loop corrections of the 2-kink. The only attempted loop calculation for non-integrable solitons was done in~\cite{Walliser1999}, where it was shown that the 2-baby-skyrmion is unstable when the corrections were included. The paper makes many approximations and our toy model could be a starting point to probe some of these.\looseness=-1

Secondly, we constructed a model with a metastable kink-antikink configuration, which we call a lav\'{\i}on. We believe this is the first time a minimal, rather than a saddle point, kink-antikink has been constructed. Perhaps other, similar configurations exist in higher-dimensional theories.

The lav\'{\i}on may also affect kink-antikink scattering, whose dynamics are highly complicated and chaotic: arbitrarily small changes in the initial velocities lead to either annihilation or bouncing of the solitons \cite{anninos1991fractal}. There is some debate over the mechanism which causes this fractal behavior: it arises in models whose kinks have bound modes \cite{campbell1983}, quasinormal modes \cite{dorey2018}, are coupled to fermions \cite{Bazeia2022} and models where only the kink-antikink pair support a normal mode when close together \cite{dorey2011}. The scattering becomes even more complicated in multi-component models \cite{alonso2020non, halavanau2012} and those with multikinks \cite{Gani2019,marjaneh2017}. The scattering structure is often attributed to the initial kinetic energy becoming temporarily ``stored'' in some other mode before settling down to the vacuum or separating infinitely. Here there is another option: the final configuration could relax into the lav\'{\i}on.

Our theories can be generalised in many ways. Introducing more vacua would allow for higher charge kink solutions. A modified sine-Gordon theory could have an arbitrary number of bound kinks. We can then ask questions about fusion and fission: ``how much energy does it cost to fission the 7-kink into a 3-kink and 4-kink? What is the lowest energy fission of the 9-kink?". These questions have direct analogs in nuclear physics and condensed matter systems.

Beyond the fundamental interest that these system represent, there is also great potential for applications. Solitons in various condensed matter systems are one of the primary objects to realize high-density memory. That motivates the search for physical systems with metastable soliton-antisoliton solutions, as they could realize dense, stable and manipulable data storage.

\subsection*{Acknowledgements}

We thank Andrzej Wereszczy\'nski for useful discussions, Anusree~N for pointing out an error in the manuscript and the referees for providing valuable feedback. CH is supported by the Carl Trygger Foundation through the grant CTS 20:25. This work is supported by the Swedish Research Council Grants 2016-06122 and 2018-03659.

\pdfbookmark[1]{References}{ref}
\LastPageEnding

\end{document}